\documentclass[aps,prl,twocolumn,nofootinbib,asmmath]{revtex4-1}
\usepackage{amssymb,esvect,amsmath,graphicx,latexsym,amsthm,slashed,eso-pic}
\usepackage[final]{pdfpages}
 \DeclareMathOperator{\kev}{keV}  \DeclareMathOperator{\gev}{GeV} \DeclareMathOperator{\tev}{TeV} \DeclareMathOperator{\cm}{cm}  \DeclareMathOperator{\g}{g}   \DeclareMathOperator{\s}{s}     
    \newcommand{\cL}{{\cal L}}   \newcommand{\cO}{{\cal O}}   
\newcommand{\ep}{\epsilon}

\newcommand{\pL}{\left(} \newcommand{\pR}{\right)} \newcommand{\bL}{\left[} \newcommand{\bR}{\right]}    
\newcommand{\beq}{\begin{equation}} \newcommand{\eeq}{\end{equation}}
\newcommand{\bea}{\begin{eqnarray}} \newcommand{\eea}{\end{eqnarray}}
\newcommand{\alg}[1]{\begin{align} \begin{split} #1 \end{split}  \end{align}}
\newcommand{\tenx}[1]{\times 10^{#1}}
\newcommand{\Eq}[1]{Eq.~(\ref{#1})}  
  
\newcommand{\Fig}[1]{Fig.~\ref{#1}}

\begin{document}

\title{Lining up the Galactic Center Gamma-Ray Excess}
\author{Samuel D. McDermott}
\affiliation{Center for Particle Astrophysics, Fermi National Accelerator Laboratory, Batavia, IL 60510}
\affiliation{Michigan Center for Theoretical Physics, Ann Arbor, MI 48109}

\begin{abstract}
Dark matter particles annihilating into Standard Model fermions may be able to explain the recent observation of a gamma-ray excess in the direction of the Galactic Center. Recently, a hidden photon model has been proposed to explain this signal. Supplementing this model with a dipole moment operator and a small dark sector mass splitting allows a large cross section to a photon line while avoiding direct detection and other constraints. Comparing the line and continuum cross sections, we find that the line is suppressed only by the relative scales and couplings. Given current constraints on this ratio, a line discovery in the near future could point to a new scale $\Lambda \sim \cO(1\tev)$, where we would expect to discover new charged particles. Moreover, such a line would also imply that dark matter can be visible in near-future direct detection experiments. %~\\ ${}$  
\hfill (FERMILAB-PUB-14-205-A-T)
\end{abstract}

\maketitle

\noindent {\bf Introduction:} As the cosmological and gravitational evidence for dark matter has grown, particle physicists have continued to seek clear indications of dark matter activity on more immediate distance- and time-scales. An excess of gamma rays observed in the region of the central Milky Way, henceforth the Galactic Center gamma-ray excess (GCGE), can be interpreted as the secondary emission from dark matter annihilations, thereby providing evidence for such a local particle dark matter population \cite{Goodenough:2009gk,Hooper:2010mq,Boyarsky:2010dr,Hooper:2011ti,Abazajian:2012pn,Gordon:2013vta,Hooper:2013rwa,Huang:2013pda,Abazajian:2014fta,Daylan:2014rsa}. A variety of authors have found a multitude of dark matter models that can accommodate the GCGE \cite{Buckley:2010ve,Logan:2010nw,Buckley:2011mm,Boucenna:2011hy,Marshall:2011mm,Zhu:2011dz,Modak:2013jya,Huang:2013apa,Okada:2013bna,Hagiwara:2013qya,Buckley:2013sca,Anchordoqui:2013pta,Berlin:2014tja,Izaguirre:2014vva,Alves:2014yha,Agrawal:2014una,Cerdeno:2014cda,Ipek:2014gua,Ghosh:2014pwa,Ko:2014gha,Boehm:2014bia,Abdullah:2014lla,Martin:2014sxa,Boehm:2014hva,Berlin:2014pya}. It is easily possible to build models that allow such a large indirect detection signal while still satisfying all constraints from direct detection, collider, and other searches.

Although explaining the GCGE through new particle physics is easy to do, verifying the dark matter origin of the GCGE will be one of the most urgent questions that particle physics will face in coming years. Other astrophysical explanations need to be fully explored, and all aspects of the theories of new physics that explain the signal must be thoroughly tested. Simply waiting to see the signal reproduced in other astrophysical regions with different systematics may take too long (and remain too systematically uncertain) to satisfy our curiosity, and there are no firm predictions for dark matter or mediator production at colliders; indeed, the new physics sector may be arbitrarily well secluded since it only needs to communicate to the Standard Model by a small amount of mass or kinetic mixing. Confirming that dark matter is responsible for the GCGE may therefore require new, observable predictions from our models.

Here we consider a hidden $U(1)$ dark matter model \cite{ArkaniHamed:2008qn,Pospelov:2008jd,Cholis:2008qq,Kang:2010mh,Finkbeiner:2007kk,Cholis:2008vb,Pospelov:2007mp,Morrissey:2009ur,Chang:2010yk,Cohen:2010kn,Meade:2009rb,Hooper:2012cw,Essig:2013goa,Berlin:2014pya} augmented by two new operators and multiple novel observables. By adding a higher dimension operator that couples dark matter to the Standard Model photon, we generate a monochromatic photon line in dark matter annihilations. Observing such a spectral feature would, in combination with current observations, give an unambiguous and experimentally robust indication that dark matter is responsible for the GCGE. Moreover, the central energy of the line would provide a clean measurement of the dark matter mass. And, intriguingly, the mass scale currently probed by the {\it Fermi} telescope and by direct detection experiments is exactly the TeV scale. Observing a line associated with the GCGE in this way would not only reveal the dark matter nature of the gamma-ray excess: the detection of a line could also provide a hint of new charged matter at accessible energies, plus an imminent direct detection signal.

~\\ \noindent {\bf A Dipole Moment:} At low scales, the dark sector Lagrangian we have in mind is
\alg{ \label{lag1}
\cL_d & = - \frac14 F_d^2+ \frac\ep2 F_d^{\mu\nu}F_{\mu\nu} - \hat g_d \bar X  \slashed A_d X - M \bar X X 
\\& + m_d^2 A_d^\mu A_{d\mu} - m \pL  X_L X_L + X_L^\dagger X_L^\dagger \pR + \frac{\hat \beta_d}\Lambda \bar X \sigma^{\mu\nu} X F_{\mu\nu},
}
where the $F$ are field strengths, $\ep$ is the kinetic mixing parameter, $A_d^\mu$ is the dark photon field, $X$ is a Dirac particle with left- and right-handed components $X_{L,R}$, and $\Lambda$ is a scale at which charged particles are integrated out. We define $\sigma^{\mu\nu}=i\bL\gamma^\mu,\gamma^\nu\bR/2$. The $U(1)_d$ is explicitly broken by the gauge boson mass and by the fermion Majorana mass. The low energy spectrum has two mass-split Majorana fermions: the mass eigenstates $X_{1,2}$ have masses $m_{1,2} \simeq | M \pm m |$, where $X_1$ is the dark matter. Crucially, the dark matter remains exactly electrically neutral and interacts with the Standard Model photon only via a nonrenormalizable transition dipole moment operator; in particular, constraints on millicharged particles do not apply. In addition, we point out that $\epsilon$ and $\hat \beta_d$ arise in principle from entirely different physical mechanisms, and a large hierarchy between these two dimensionless parameters is possible. Variations on dark matter with elastic or inelastic magnetic moments have been considered in other contexts \cite{Sigurdson:2004zp,Chang:2010en,Tulin:2012uq}. The Lagrangian in \Eq{lag1} differs from the hidden photon model of \cite{Berlin:2014pya} only by the two final terms in \Eq{lag1}.

These new terms are of great phenomenological interest. Without them, the model does not provide tree-level dark matter-photon vertices, and, like all other models proposed for the GCGE, can only deliver a photon line by closing the final state Standard Model fermion loop. The cross section for a photon line resulting from such loop processes is expected to be
\beq
\langle \sigma v \rangle_{\gamma \gamma {\rm,\,loop}} \sim \langle \sigma v \rangle_{ff} \times e^4/16\pi^2 \sim 10^{-31} \cm^3/\s,
\eeq
i.e., many orders of magnitude lower than the cross section for dark matter annihilation to fermions, and more than two orders of magnitude below the current {\it Fermi} bounds \cite{Ackermann:2013uma}. Such a low cross section is unlikely to be probed by near future gamma-ray telescopes. This is a typical feature of models that have no tree-level interactions between the dark matter and the photon (however, see \cite{Tulin:2012uq} for important exceptions). In order to produce monochromatic photons in a non-negligible portion of dark matter annihilations, we must therefore add in a new operator by hand that allows the photon to couple to the dark current at tree level. As long as the dark matter remains electrically neutral, gauge invariance requires that at low energies such a coupling manifest as the final term of \Eq{lag1} -- this is a dipole moment operator. Such a higher dimension operator can be generated by integrating out charged particles; the dimensionful suppression scale of the operator is generally the scale at which these new particles can go on shell.

\begin{figure}[t]
\begin{center}
\includegraphics{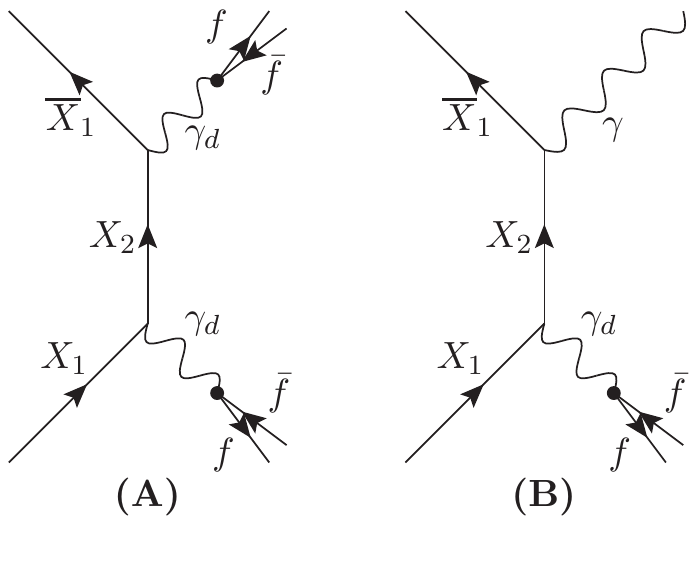}
\caption{Annihilation through diagram {\bf(A)} sets the relic density in the early universe and accounts for the GCGE. Annihilation through diagram {\bf(B)} gives a photon line (as well as a subdominant amount of continuum photons). Annihilation to two on-shell photons is suppressed relative to {\bf(B)}.}
\label{dip-ann}
\end{center}
\end{figure}

In the mass eigenbasis, \Eq{lag1} contains a dark flavor changing neutral current,
\beq \label{LdFC}
\cL_d \supset g_d \bar X_1 \slashed A_d X_2 +{\rm h.c},
\eeq
where $g_d \equiv \hat g_d \times m/M$. As long as $m_d \leq m_1$, this allows the annihilation $X_1 \bar X_1 \to \gamma_d \gamma_d$, as shown in panel {\bf(A)} of \Fig{dip-ann}. The GCGE continuum photons and the relic density simply come about from annihilations to $\gamma_d$ in exactly the same way as in the hidden sector model of \cite{Berlin:2014pya}: pairs of dark matter particles annihilate to the slightly lighter vector $\gamma_d$, which propagates over a macroscopic distance before decaying to Standard Model fermions from the kinetic mixing of the $\gamma_d$ with the Standard Model photon. In the early universe, this process remains in equilibrium until $n_X(T) \langle \sigma v\rangle_{\gamma_d \gamma_d} $ falls below the Hubble rate, leaving a relic density of $X_1$ and $\bar X_1$ particles. The annihilation cross section required to attain the cosmological abundance after this freeze out process is the same cross section as required to match the flux of photons from the GCGE.

Aside from the dark current in \Eq{LdFC}, the interaction terms of \Eq{lag1} also include a transition magnetic dipole moment in the mass basis,
\beq \label{Ldip}
\cL_d \supset \frac{\beta_d}\Lambda  \bar X_1 \sigma^{\mu\nu} X_2 F_{\mu\nu} +{\rm h.c},
\eeq
where $\beta_d \equiv \hat \beta_d \times m/M.$ Of particular interest for this work, and in contrast to prior studies, this term allows annihilations that include monochromatic photons, as shown in panel {\bf(B)} of \Fig{dip-ann}. These photons will have energy $E_\gamma = (4m_1^2-m_d^2)/4m_1 \simeq 3m_1/4$, which in addition to more refined GCGE spectra will allow a clean determination of the dark matter mass.

The annihilation cross sections shown in \Fig{dip-ann} both have $s$-wave terms. In the low-velocity limit and taking all dark sector masses to be set by the common mass $m_X \equiv m_1 \sim m_d \sim m_2$, we find
\alg{ \label{xSecs}
\langle \sigma v \rangle \underset{v\to0}{\simeq} \left\{\begin{array}{lr} g_d^4 p_{d,{\bf A}}^3/4\pi m_X^5 & ( X_1 \bar X_1 \to \gamma_d \gamma_d ) \\ 
8\beta_d^2 g_d^2 p_{d,{\bf B}}^3/9\pi m_X^3\Lambda^2 & ( X_1 \bar X_1 \to \gamma \gamma_d) \\ 
\beta_d^4m_X^2/4 \pi\Lambda^4 & ( X_1 \bar X_1 \to \gamma \gamma ) \end{array} \right.
}
where $p_{d,{\bf A}} =  \sqrt{m_1^2-m_d^2}$ is the momentum of the outgoing dark photons for the annihilation shown in \Fig{dip-ann} {\bf(A)} and $p_{d,{\bf B}} = (4m_1^2-m_d^2)/4m_1$ is the momentum of the outgoing dark photon for the annihilation shown in \Fig{dip-ann} {\bf(B)}. For completeness, we have also calculated the cross section to two photons. The expressions in \Eq{xSecs} are valid at the 10\% level for generic mass splittings less than 20\%, but we use the exact expressions in all numerical work.

\begin{figure}[t]
\begin{center}
\includegraphics[width=.46\textwidth]{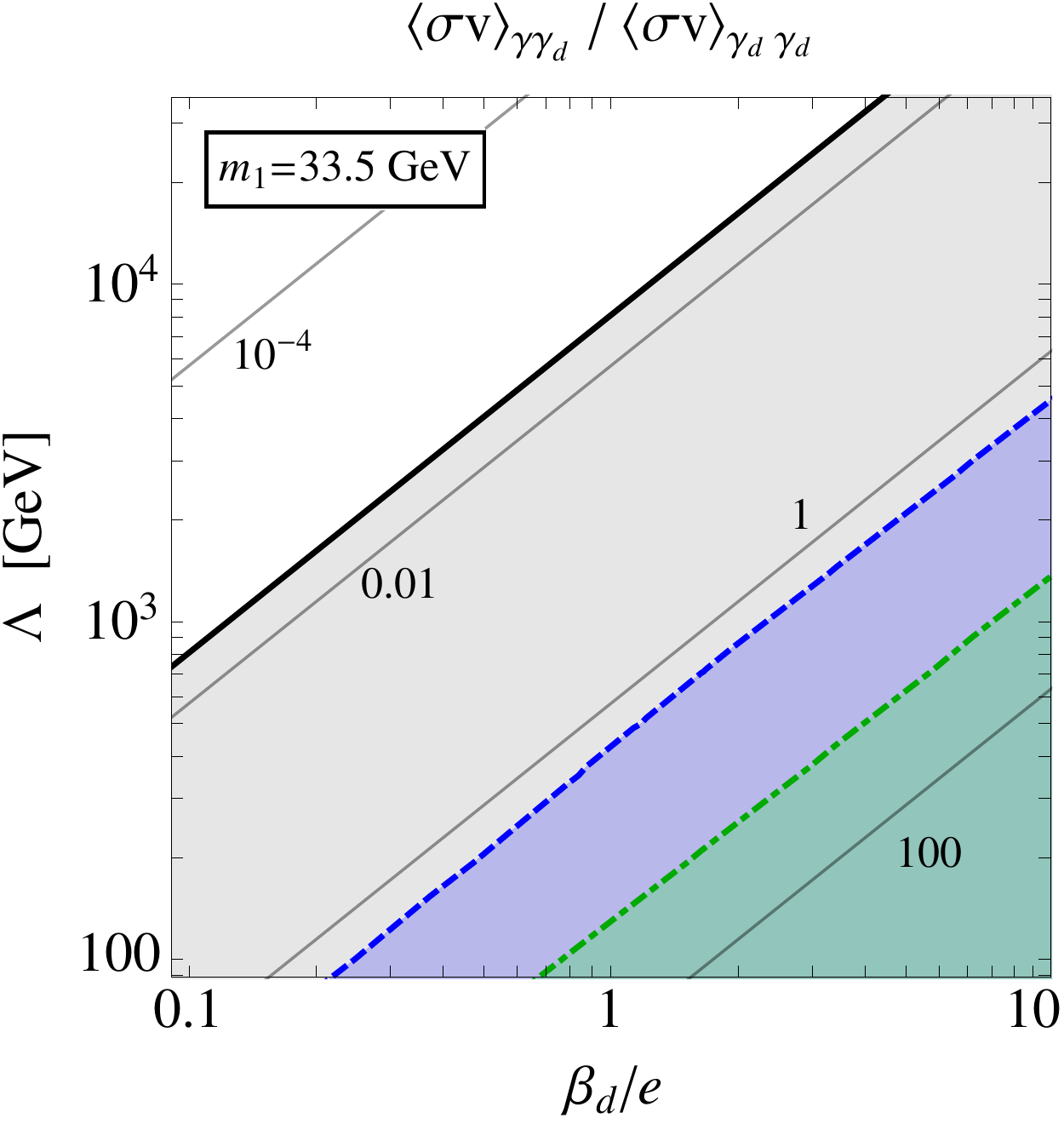}
\caption{Contours of $\langle \sigma v \rangle_{\gamma \gamma_d}/\langle \sigma v \rangle_{\gamma_d \gamma_d}$ as a function of the coupling $\beta_d$ and the scale $\Lambda$. The gray shaded region below the solid line is ruled out by {\it Fermi} line searches. The blue shaded region below the dashed line is ruled out by LUX constraints; the green region below the dot-dashed line is ruled out by electroweak measurements. We ensure a good fit to the relic density and the GCGE by taking $g_d=0.1$, $m_1=33.5\gev$, and fixing $m_2 : m_1 : m_d = 1.1 : 1 : 0.9$.}
\label{rat-mx}
\end{center}
\end{figure}

From \Eq{xSecs}, we see that each on-shell photon suppresses the cross section by roughly $\beta_d^2 m_X^2 / g_d^2 \Lambda^2$. {\it Fermi} bounds on the cross section to a photon line are currently $\langle \sigma v \rangle_{\gamma \gamma_d} \lesssim 10^{-28}\cm^3/\s$ \cite{Ackermann:2013uma}\footnote{{\it Fermi} searches constrain $\langle \sigma v \rangle_{\gamma \gamma}$, while our model produces a single photon, so the limits from \cite{Ackermann:2013uma} are weakened by a factor of two. The number quoted assumes an NFW dark matter profile and is strengthened or weakened by a factor of a few for different dark matter profiles. However, the bounds are highly sensitive to the dark matter mass, so we take a representative bound.}, compared to the normalization required for the GCGE, $\langle \sigma v \rangle_{\gamma_d \gamma_d} \sim 2 \tenx{-26}\cm^3/\s$ \cite{Daylan:2014rsa,Berlin:2014pya}. Hence, if there is no kinematic suppression, the approximations in \Eq{xSecs} indicate
\beq \label{Lambda-est}
\frac{4 \beta_d^2 m_X^2 }{ g_d^2 \Lambda^2} \lesssim 5\tenx{-3} \implies \Lambda \gtrsim \tev \times \frac{\beta_d}{g_d} \times \frac{m_X}{30\gev}.
\eeq
Finding a photon line associated with the GCGE in upcoming {\it Fermi} data would thus point towards new charged particles at the TeV scale, unless there is a large hierarchy in $\beta_d/g_d$. We plot the ratio $\langle \sigma v \rangle_{\gamma \gamma_d}/\langle \sigma v \rangle_{\gamma_d \gamma_d}$, including the exact expressions for the annihilation cross sections, in \Fig{rat-mx}. We set the masses proportionally as $m_2 : m_1 : m_d = 1.1 : 1 : 0.9$, and we fix $m_1=33.5\gev$. This sets the dark photon mass $m_d \simeq 30\gev$; these masses can explain the GCGE at the 1$\sigma$ level \cite{Berlin:2014pya}. The mass ratio here also gives a $X_1-X_2$ mass splitting of a few GeV, which is relevant for the remaining bounds.

~\\ \noindent {\bf Additional Phenomenological Implications:} We list some bounds that can constrain the model of \Eq{lag1}.

{\it Collider bounds:} Because of the small couplings to Standard Model fermions (all of which are suppressed by $\ep$ or $\Lambda$), collider searches for $X_1$ or $\gamma_d$ should be weak. However, $\Lambda$ indicates a mass scale where new particles which are charged under $U(1)_{\rm EM}$ can go on shell, so bounds on heavy stable charged particles provide a different test of the theory that effectively places bounds on $\Lambda$ alone (rather than the ratio $\beta/\Lambda$).

As an example of the bounds on charged particles, we note that LEP requires that the chargino $\chi^\pm$ of the MSSM satisfy $m_{\chi^\pm} \geq 103.5 \gev$ \cite{LEPchi}, while ATLAS \cite{Aad:2014vma} and CMS \cite{Khachatryan:2014qwa} searches for charged SUSY particles are generally in the several hundred GeV range. These searches rely on model dependent final state signatures, so we do not make model independent assertions here. It suffices to say that new electromagnetically charged particles with masses around the TeV scale are both currently acceptable and potentially discoverable at the LHC. That the TeV scale falls out of the current {\it Fermi} line bounds is an exciting prediction of our model.

{\it Direct Detection:} The maximum energy deposition possible in terrestrial direct detection experiments is $E_{\rm kin,max}^{\rm DD} \simeq m_{\rm DM} v_{\rm esc}^2 /2 \simeq 50 \kev \times m_{\rm DM}/30\gev$. If the mass splitting $2m$ exceeds this scale, there is no tree-level scattering at direct detection experiments, neither through the higher dimension dipole operator, nor through the renormalizable scattering from $\gamma_d - \gamma$ mixing. The scattering at direct detection experiments will only enter at loop level, for which we estimate (assuming a common dark sector mass $m_X$)
\beq
\sigma_{\rm SI} \sim \left\{ \begin{array}{lr} g_d^4 \ep^4 e^4 m_p^2/16\pi^2 m_X^4 & {\rm(via~\gamma}_d{\rm~exchange})
\\ \beta_d^4 e^4 m_p^2/16\pi^2\Lambda^4 & {\rm(via~\gamma~exchange}) \end{array} \right. .
\eeq
Using the LUX \cite{Akerib:2013tjd} constraints\footnote{We have rescaled to account for the fact that scattering is only off protons and not off the entire nucleus \cite{Berlin:2014pya}.} on 33.5 GeV dark matter, $\sigma_{\rm SI} \lesssim 4.6 \tenx{-45}\cm^2$, we find $\ep \lesssim 2\tenx{-2}/g_d$ (about two orders of magnitude weaker than found for the elastic scattering case \cite{Berlin:2014pya}) and $\Lambda \gtrsim 1.4\tev \times \beta_d$, the latter of which we show in \Fig{rat-mx}.\footnote{The mixed case gives $\Lambda \gtrsim 140 \tev \times (\beta_d \ep g_d).$ Though this is competitive, it relies on $\ep$, which may be very small.}

Because the direct detection constraint on the dipole moment is currently only an order of magnitude weaker than the line search bound and will rapidly strengthen, a line observation consistent with our model gives an expectation for observation at next-generation direct detection experiments. The direct detection cross section indicated by the current {\it Fermi} sensitivity is of order $\sigma_{\rm SI} \sim \cO(10^{-49}\cm^2)$, which is just above the ``neutrino floor" for $m_1=33.5\gev$.

{\it  Electroweak Precision:} The dipole interaction can affect the precision measurement of well-understood electroweak observables. Because these effects come in at loop level, they are insensitive to the mass splitting that suppresses the rate in direct detection experiments. The most relevant observations are the muon magnetic moment, the perturbativity of the model at the $Z$ pole, and the running of the fine structure constant as measured by the ratio of the $W$ mass squared to the Fermi constant \cite{Sigurdson:2004zp}, listed in increasing order of severity. The running of $\alpha$ requires $ \Lambda \gtrsim 440 \gev \times \beta_d $ \cite{Sigurdson:2004zp}, which confirms the intuition that new electrically charged particles must be more massive than a TeV. We show this bound in \Fig{rat-mx}.

 {\it Self-Interactions:} Analogous to direct detection, there are two channels for self-interaction in this model, and, due to the inelastic nature of the low energy Lagrangian, self scattering that remains in the lower mass state occurs at loop level. In the limit of degenerate masses, the leading order scattering cross section may be estimated by dimensional analysis as
\beq
\sigma_{\rm self} \sim \left\{ \begin{array}{lr}  g_d^8/16\pi^2 m_X^2 & ({\rm via~ \gamma}_d~{\rm exchange}) \\ \beta_d^8 m_X^6  /16\pi^2 \Lambda^8 &  ({\rm via~ \gamma~ exchange}) \end{array} \right. .
\eeq
Considerations of the Bullet Cluster require $\sigma_{\rm self}/m_X \lesssim 1\cm^2/\g$ \cite{Clowe:2006eq}. At $m_1=33.5\gev$ we find $g_d \lesssim 20$ and $\beta_d \lesssim 600$; these constraints are less restrictive than the requirement of perturbativity.

{\it Cosmology:} Since the lifetime for $X_2 \to X_1 \gamma$ decay is very short and occurs well before BBN, the strongest bounds from cosmology in this model are derived from requiring that the dark matter not couple too strongly to matter in the epoch of recombination. The leading bounds \cite{Dvorkin:2013cea} are weakened by a loop since typical momentum transfers in the CMB epoch will fail to breach the inelastic scattering threshold, in analogy with direct detection. We find essentially nonexistent model bounds.

{\it Magnetic Field Interactions:} Adding a magnetic interaction to the dark sector may seem problematic because there are strong magnetic fields in the galactic center, in the form of SNe remnants, a large plane-parallel component, and turbulent eddies, with an overall magnitude on large scales of order $10-100\,\mu$G \cite{Haverkorn:2014jka}. However, the potential energy for aligning along these field lines, $H_B \sim \beta_d B/\Lambda$, is still many orders of magnitude lower than the kinetic energy of a typical dark matter particle near the center of the galaxy, and magnetic effects should be unimportant for the gross features of the signal.

~\\ \noindent {\bf Model Building:} Although building a UV complete model that gives rise to the Lagrangian of \Eq{lag1} at low scales is beyond the scope of this work, we make a few remarks here. In addition to a new charged fermion $X_\pm$ with mass $m_\pm \simeq \Lambda \gtrsim \tev$, we need a new Higgs field whose vev spontaneously breaks the symmetries of the UV theory and provides the dark sector masses. This dark Higgs will need a charged component $H_\pm$ to couple $X$ with $X_\pm$. Because these particles must be charged under $U(1)_{\rm EM}$, they must have electroweak quantum numbers. Finally, the neutral components of the dark sector Higgs must mix very weakly with the Higgs of the Standard Model to avoid large direct detection rates \cite{Berlin:2014pya}. This list of model building requirements is by no means trivial, but it should be possible to satisfy.

Even in the absence of a UV-complete theory, we can estimate the coupling $\beta_d$. Since it should arise at one loop when $X$ splits into a $X_\pm$ and $H_\mp$, we estimate
\beq \label{beta-est}
\frac{\lambda_d^4 e_X^2}{16\pi^2 m_\pm^2} \sim \frac{\hat \beta_d^2}{\Lambda^2} \implies \beta_d \sim \frac mM \frac\Lambda{m_\pm} \frac{\lambda_d^2 e_X }{4\pi } ,
\eeq
where $\lambda_d$ is the $X-X_\pm-H_\mp$ coupling, $e_X$ is the electromagnetic charge of the $X_\pm$, and $m,M$ are the masses in \Eq{lag1}. If $m_\pm\simeq\Lambda\sim\tev$ (see \Eq{Lambda-est}) and $m \sim M/10$, then in order for $\beta_d\sim g_d \sim \cO(0.1)$ we see that $\lambda_d$ must be $\sim \cO(1)$, and may even have to be near strong coupling. However,  because we have decoupled the interaction with the photon from the physics that controls the kinetic mixing, the parameter $\beta_d$ may be large even if $\epsilon$ is relatively small. The rough estimate in \Eq{beta-est} indicates that it could even be interesting to consider the consequences of the low energy $X$ being a strongly coupled composite particle, where $\Lambda$ is now seen as some new QCD scale.

~\\ \noindent {\bf Conclusions:} We have shown that a photon line induced by a transition magnetic dipole moment can be observed in {\it Fermi} line searches while retaining the phenomenological successes of the hidden photon model \cite{Berlin:2014pya} and avoiding direct and indirect constraints. Should a line be observed with additional {\it Fermi} data, we will have unambiguous support for a dark matter explanation of the GCGE, not to mention a sharp kinematic measurement of the dark matter mass. In the event of such an observation, the TeV scale falls out ``for free," and our simple low energy model has exciting implications for LHC physics and near-future direct detection experiments.

~\\ \noindent {\bf Acknowledgments:} We thank Prateek Agrawal, Dan Hooper, and Tongyan Lin for collaboration on early stages of related work. We thank Kfir Blum for correspondence. We additionally thank Prateek, Dan, Tongyan, Asher Berlin, and Kathryn Zurek for comments on the draft. We acknowledge Stony Brook University, Brookhaven National Lab, the Dark Interactions Workshop, and TeVPA/IDM 2014, where this work was completed.  The author is supported by the Fermilab Fellowship in Theoretical Physics. Fermilab is operated by Fermi Research Alliance, LLC, under Contract No.~DE-AC02-07CH11359 with the US Department of Energy.

\bibliography{hiddengev}

\end{document}